\documentclass[11pt]{article}
\usepackage{amsmath,amssymb,bm}
\usepackage{graphics}
\usepackage{graphicx}
\usepackage{amssymb}
\usepackage{epstopdf}
\usepackage[utf8]{inputenc}

\usepackage{mystyle}
\usepackage{cite,./mcite}
\usepackage{authblk}
\def\pythia{{\sc Pythia}}

\title{Recent \pythia~8 developments: Hard diffraction,\\ Colour reconnection and $\gamma\gamma$ collisions}

\author{Ilkka Helenius}
\author{Jesper R.~Christiansen}
\author{Christine O.~Rasmussen}
{\tiny
\affil{Department of Astronomy and Theoretical Physics, Lund University, S\"{o}lvegatan 14A, SE-223 62 Lund, Sweden}
} 

\begin{document}

\maketitle 

\vspace{-20\baselineskip}
\mbox{}\hfill LU TP 16-01\\
\mbox{}\hfill MCnet-16-09\\
\mbox{}\hfill April 2016\\
\vspace{15\baselineskip}

\section{Introduction}
An overview of recent developments in \pythia~8 is given. First the new hard diffraction model, which is implemented as a part of the multiparton interactions (MPI) framework, is discussed. Then the new colour reconnection model, which includes beyond leading colour effects that can become important when MPI are present, is briefly reviewed. As a last topic an introduction is given to our implementation of photon-photon collisions. In particular photon PDFs, required modifications for the initial state radiation algorithm and beam remnant handling with photon beams is discussed.

\section{Hard Diffraction}

A model for soft diffraction has long been available in \textsc{Pythia}~8 and earlier versions \cite{Sjostrand:2006za,Sjostrand:2014zea}. This model allows for $2\rightarrow2$ QCD processes at all $p_{\perp}$ scales, but is primarily intended for lower values of $p_{\perp}$. For truly hard diffractive processes, the new model for hard diffraction \cite{Rasmussen:2015qgr} was developed, not only for high-$p_{\perp}$ jets, but also allowing for $W^{\pm}, Z^0, H$ etc.

The model is based on the assumption that the proton PDF can be split into a diffractive and a nondiffractive part,
\begin{align}
f_{i/\mathrm{p}}(x,Q^2) &= f_{i/\mathrm{p}}^{\mathrm{ND}}(x,Q^2) +
f_{i/\mathrm{p}}^{\mathrm{D}}(x,Q^2) \,\,\, \mathrm{with}\\
f_{i/\mathrm{p}}^{\mathrm{D}}(x, Q^2) &=  \int_x^1  \frac{\mathrm{d} x_{\mathbb{P}}}{x_{\mathbb{P}}} \,  
  f_{\mathbb{P}/\mathrm{p}}(x_{\mathbb{P}}) \, f_{i/\mathbb{P}} \left( 
  \frac{x}{x_{\mathbb{P}}}, Q^2 \right) ~,
\end{align}
with $x$ and $Q^2$ the momentum fraction and virtuality of parton $i$, $x_{\mathbb{P}}$ and $t$ the momentum fraction and virtuality of the Pomeron, and where the Pomeron flux $f_{\mathbb{P}/\mathrm{p}}(x_{\mathbb{P}}) = \int_{t_{\mathrm{min}}}^{t_{\mathrm{max}}}\mathrm{d} t \, f_{\mathbb{P}/\mathrm{p}}(x_{\mathbb{P}}, t)$, as $t$ for the most part is not needed. The tentative probability for side $A$ to be diffractive is then given by the ratio of diffractive to inclusive PDFs,
\begin{align}
\label{Eq:Prob}
\mathcal{P}_A^{\mathrm{D}} &= \frac{f_{i/B}^{\mathrm{D}}(x_B, Q^2)}
{f_{i/B}(x_B, Q^2)}~~~\mathrm{for}~~~AB \to XB~
\end{align}
with a similar equation for side $B$. The model further implements a dynamical rapidity gap survival, cf.\ Fig.~\ref{Fig:DynGapSur}. On an event-by-event basis the possibility for additional MPIs in the $\mathrm{pp}$ system is evaluated, and if no such MPIs are found, then the event is diffractive. Only then is the $\mathbb{P}\mathrm{p}$ system set up and a full evolution is
performed in this subsystem, along with the hadronization of the colour strings in the event. 

%\begin{figure}
%\centering
%\includegraphics[width=0.44\textwidth]{singleDiffraction.pdf}
%\caption{\label{Fig:DynGapSur} The dynamical gap survival implemented 
%in \textsc{Pythia}~8. The model does not allow for MPIs in the
%$\mathrm{pp}$ system, but allows for additional MPIs in the
%$\mathbb{P}\mathrm{p}$ system.}
%\end{figure}

The dynamical gap survival introduces an additional suppression of the diffractive events, such that the total probability for hard diffraction drops from $\sim10$\% to $\sim1$\%, exact numbers depending on parametrization of $\mathbb{P}$ flux and PDF as well as the free
parameters of the MPI framework in \pythia~8. The new model is compatible with suppression factors measured at the LHC.

\section{Colour Reconnection}

The colour reconnection (CR) refers to a phenomenon where the colour strings formed during the hard process and parton shower generation can reconnect and form a different configuration prior to hadronization. A new model for CR in \pythia~8 was introduced in Ref.~\cite{Christiansen:2015yqa}. The model is based on three main principles:
use of SU(3) colour rules to determine if the strings are colour compatible, %so that the reconnection is possible and what is the probability for the configuration.
a simplistic space-time picture to check whether the two colour strings can be causally connected, and
minimization of string length measure $\lambda$ to find which colour configurations are preferred.
%\begin{itemize}
%\item Use of SU(3) colour rules to determine if the strings are colour compatible.% and what is the probability for the configuration.
%\item A simplistic space-time picture to check whether the two colour strings can be causally connected.
%\item Minimization of string length measure $\lambda$ to find which colour configurations are preferred.
%\end{itemize}
In addition to simple colour strings between a quark and an antiquark, the new model includes also junction structures which can connect three (anti)quarks together. As the junction structures are related to baryons, an enhanced baryon production can be expected.
%, inclusion of these are expected to lead to an enhancement in baryon production. 

Natural observables to study the CR effects are baryon-to-meson ratios. The parameters related to the new CR model was fixed using the CMS data for the rapidity dependence of the $\Lambda/K_S^0$ ratio, while keeping the rate in $\mathrm{e}^+\mathrm{e}^-$ collisions unaffected. Without further tuning also the $p_{\perp}$ dependence of this ratio can be described more accurately than with the old \pythia~model, although there still exists some discrepancy for $p_{\perp}>5~\mathrm{GeV/c}$.

\begin{figure}[t]
\begin{minipage}[t]{0.445\linewidth}
\centering
\includegraphics[trim = 0pt -12pt 0pt 0pt, clip, width=0.95\textwidth]{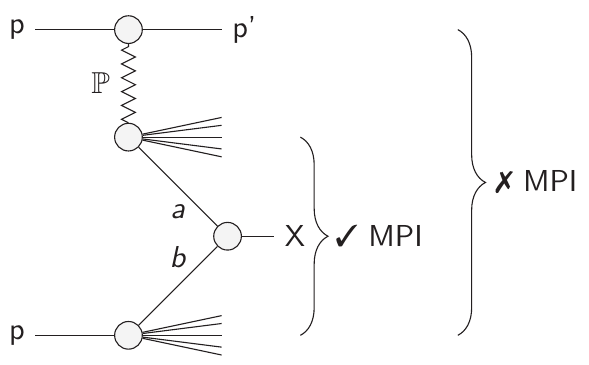}
\caption{The dynamical gap survival implemented 
in \textsc{Pythia}~8. The model does not allow for MPIs in the
$\mathrm{pp}$ system, but allows for additional MPIs in the
$\mathbb{P}\mathrm{p}$ system.}
\label{Fig:DynGapSur}
\end{minipage}
\hspace{0.01\linewidth}
\begin{minipage}[t]{0.535\linewidth}
\centering
\includegraphics[width=\textwidth]{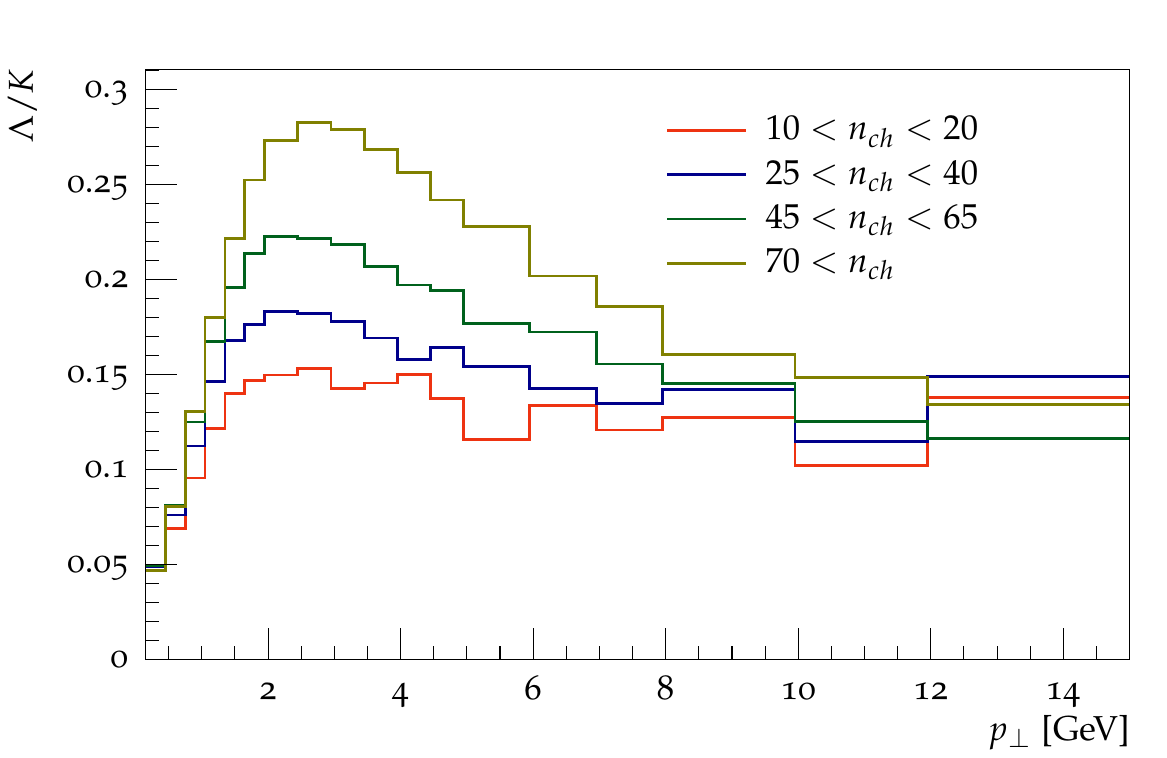}
\caption{Transverse momentum dependence of the $\Lambda/K$ ratio for different multiplicities in p+p collisions.}% at $\sqrt{s} = 7.0\,\mathrm{TeV}$.}
\label{fig:lambdaKmult}
\end{minipage}
\end{figure}

A novel way to constrain different CR models is to study the multiplicity dependence of the baryon-to-meson ratios \cite{Bierlich:2015rha}. The old \pythia~model is more or less flat with multiplicity but -- due to the junction structures -- the new model predicts an enhancement for baryon production with higher multiplicities. A very interesting observable is the $p_{\perp}$ dependence of $\Lambda/K$ ratio in different multiplicities, shown in Fig.~\ref{fig:lambdaKmult}. In the lowest multiplicity bin the ratio is flat above $p_{\perp} = 2~\mathrm{GeV/c}$, but with higher multiplicities there is a growing enhancement at intermediate $p_{\perp}$, flattening again at $p_{\perp}\gtrsim 8~\mathrm{GeV/c}$. Also, the peak of the ratio shifts towards higher values of $p_{\perp}$ with higher multiplicities -- a behaviour that is typically connected to collective flow in heavy-ion collisions. To sort out the physical origin of these kind of observations, further studies are of great interest.

\section{Photon-photon Collisions}

The $\gamma\gamma$ collisions give access to several interesting hard processes and provide an additional test of QCD factorization. A further motivation to study these collisions is provided by the future $\mathrm{e}^+\mathrm{e}^-$ experiments, where the $\gamma\gamma$ interactions will generate background processes that influence to the physics potential of future measurements.

An option to simulate $\gamma\gamma$ collisions with \pythia~8 have been added with the 8.215 release. The new implementation is not just a repetition of the framework included in \pythia~6 \cite{Sjostrand:2006za} but a new -- more robust -- machinery has been under development \cite{HeleniusSjostrand}. Currently hard processes with parton showers and hadronization of resolved $\gamma\gamma$ interactions can be generated for real photons, but MPIs and soft processes are not included. Here the key differences to the hadronic collisions are briefly described.

\subsection{Evolution equations and parton showers}

The partonic structure of resolved photons is described by the PDFs in a similar manner as for hadrons. The scale evolution of the photon PDFs is given by
\begin{equation}
\frac{\mathrm{\partial} f^{\gamma}_i(x,Q^2)}{\mathrm{\partial}\mathrm{log}(Q^2)} = \frac{\alpha_{\rm EM}}{2\pi}e_i^2 P_{i\gamma}(x) + \frac{\alpha_s(Q^2)}{2\pi} \sum_j \int_x^1\frac{\mathrm{d}z}{z}\, P_{ij}(z)\, f_j(x/z,Q^2),
\end{equation}
where the first term on the right hand side corresponds to the $\gamma \rightarrow q \bar{q}$ splitting of the beam photon and the second to the usual partonic splittings. Due to the first term, the scale evolution increases the quark PDFs over the whole $x$ region and thus increases the large-$x$ quark PDFs in photons compared to quark PDFs in hadrons.% In our studies we have used photon PDFs from CJKL analysis \cite{Cornet:2002iy}.

As the parton shower is generated using the DGLAP equations, the \pythia~algorithm \cite{Sjostrand:2004ef} must be extended accordingly by including a possibility to find the original beam photon during the generation of the initial state radiation (ISR). If the beam photon is found, no further ISR evolution is allowed for the beam.% The final state radiation can be generated using the same algorithm as for hadrons as here only splittings of individual partons are considered and the origin of the partons does not play a role.

\subsection{Beam remnants}

The beam remnants describe the leftovers of the beam particle after the interaction, for details of \pythia~8 implementation see Ref.~\cite{Sjostrand:2004pf}. Unlike in the case of protons, the valence content of photons is not known beforehand. %but it can be a $q \bar{q}$ pair of any flavour. 
To fix for the valence content the PDFs are decomposed into sea and valence contributions. These are then used to decide whether the parton taken from the beam was a valence parton or not. If it was a valence quark, the remnant is simply the corresponding (anti)quark and if not, the valence content is sampled according to the PDFs. After the valence content is fixed, a minimal number of partons are added to make sure that the flavour and colour are preserved in an event. If the ISR algorithm ends up at the beam photon, there is no need to add any remnant partons for the given beam.

As the quark PDFs for photons have a large contribution also from large values of $x$, it may happen that there is no room to add the beam remnants with non-zero masses. Also the parton shower can lead to a situation where the remnants can not be constructed due to the lack of invariant mass available for the remnants. These cases are rejected during the event generation to make sure that only events with appropriate remnants are generated.

\section*{Acknowledgements}
Work has been supported by the MCnetITN FP7 Marie Curie Initial Training Network, contract PITN-GA-2012-315877 and has received funding from the European Research Council (ERC) under the European Union's Horizon 2020 research and innovation programme (grant agreement No 668679).

\bibliographystyle{MPI2015}
\bibliography{references}

\end{document}